\documentclass[letterpaper,10pt]{article}%
\pdfoutput=1
\usepackage{osameet2}
\usepackage{amsmath,amssymb}
\usepackage{amsmath}%
\usepackage{amsfonts}%
\usepackage{amssymb}%
\usepackage{graphicx}
\usepackage{subcaption}
\providecommand{\U}[1]{\protect\rule{.1in}{.1in}}
\newtheorem{theorem}{Theorem}

\begin{document}

\title{A Practical Layered Model for Flexible-grid Optical Networks to Reduce the Routing Complexity}
\author{Ay\c{s}eg\"{u}l Yay{\i}ml{\i}}
\address{Istanbul Technical University, Istanbul, Turkey, Email: gencata@itu.edu.tr}

\begin{abstract}
This article proposes a new layered model to represent the spectrum assignment on flexible-grid optical 
networks. This model can reduce the time-complexity of existing routing and spectrum 
assignment methods by providing a data structure that captures the current spectrum usage of the
network links. 
\end{abstract}

\section{Introduction}
Flexible-grid optical networks allow the operators to use the fiber's spectrum more 
freely, based on a much finer grid (e.g., 12.5 GHz spacing).  
Unlike the WDM technology where each connection is routed using a wavelength, 
in flexible-grid networks, an optical channel can span several 
consecutive slots (smallest usable spectrum portion) to accommodate the required bandwidth. This flexibility 
brings together a difficulty in modeling the spectrum usage. Traditionally, each wavelength in a WDM network can 
be visually represented by creating a copy of the physical network topology. 
However, this representation is not valid 
in flexible-grid context, since it cannot represent the interdependencies between the neighboring slots.

Here, we present a new layered structure that can model the network state in terms of connection
routing and slot assignment. First, we introduce the model on a single link. 
Then we discuss how to use this representation in a more general network setting, where the nodes 
are connected in a general mesh topology. We also present an example usage of the new model by a 
simple routing and slot assignment (RSA) heuristic, and elaborate on how to update resource usage information
in dynamic traffic conditions.

\section{Single link model}
A fiber's usable spectrum is partitioned into a predefined number of slots, denoted by $T$. 
Connections of different bandwidth requests (e.g., 40 Gbps, 100 Gbps, 400 Gbps) can be allocated
on one or several consecutive slots according to the modulation format of the optical signal. 
In the literature, the total number of slots $T$ on a fiber is typically assumed to be between 32 and 400, 
whereas the required number of slots to accommodate an incoming connection request is assumed to be
between 1 and 32. Also, previous studies suggest two request patterns:
a uniform pattern where any number of slots between $[1,k]$ (k is 4, 10, 16, or 30 in referred papers) 
can be used\cite{chen,fun,chris} for a connection;
and a power-of-two pattern where the required number of slots are assumed to be a power of 
two\cite{almeida,castro} (the largest request requires 8, 16, or 32 slots in referred papers). 
The model developed here is shown on these two models,
however the proposed model is general in the sense that it can be used to model any other reuqest pattern. 
we should note that including or omitting the guard bands (spectrum slots left unused in between the neighboring 
bands carrying connections) does not affect the design of the model.

\subsection{Uniform requests}
\begin{figure}[t]
\centering
\begin{subfigure}{0.49\textwidth}
\centering
\includegraphics[width=7cm]{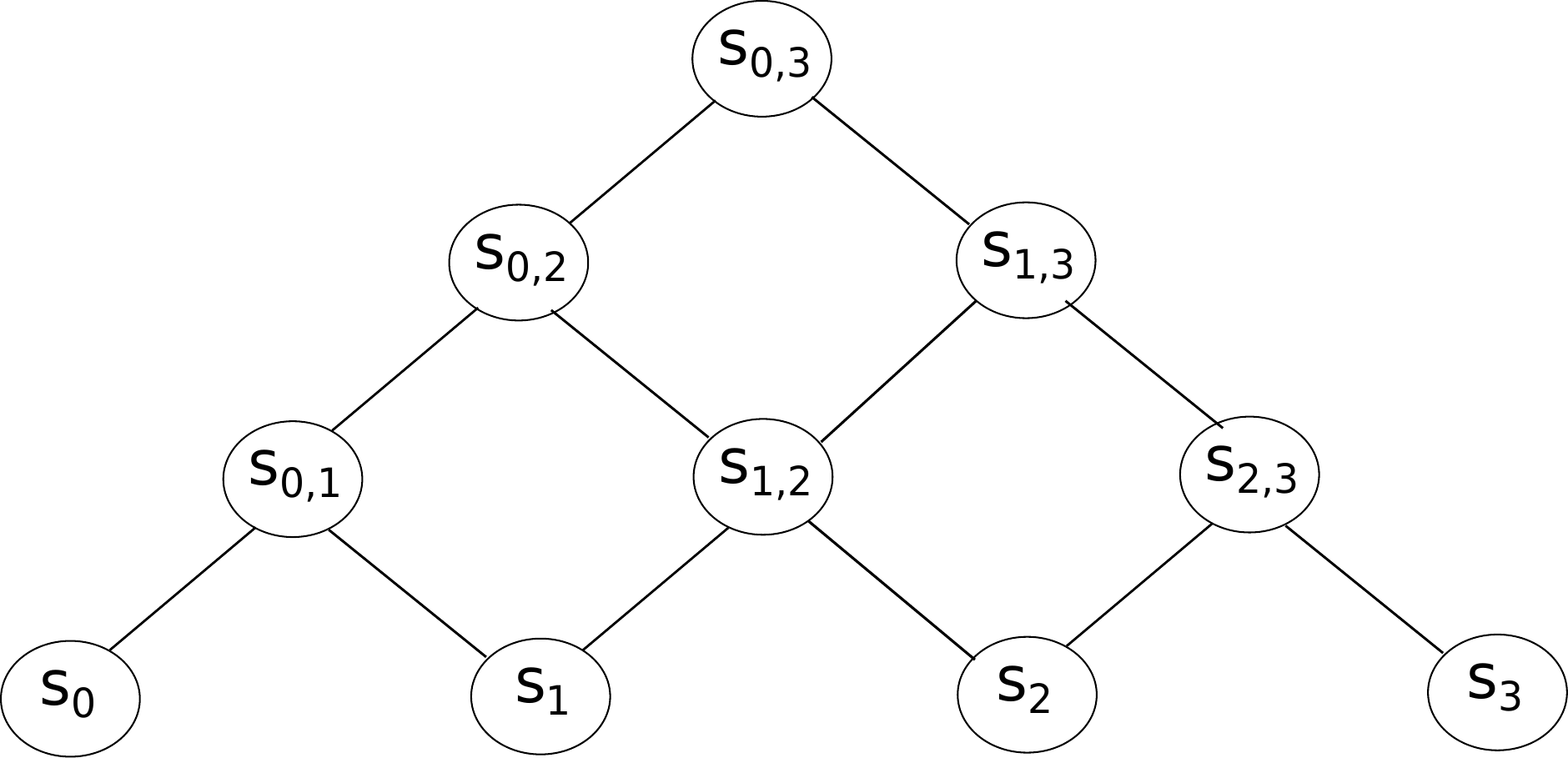}
\caption{Model for uniform requests. $T=4, k=4$}
\label{Hasse:uniform1}
\end{subfigure}
\begin{subfigure}{0.49\textwidth}
\centering
\includegraphics[width=7cm]{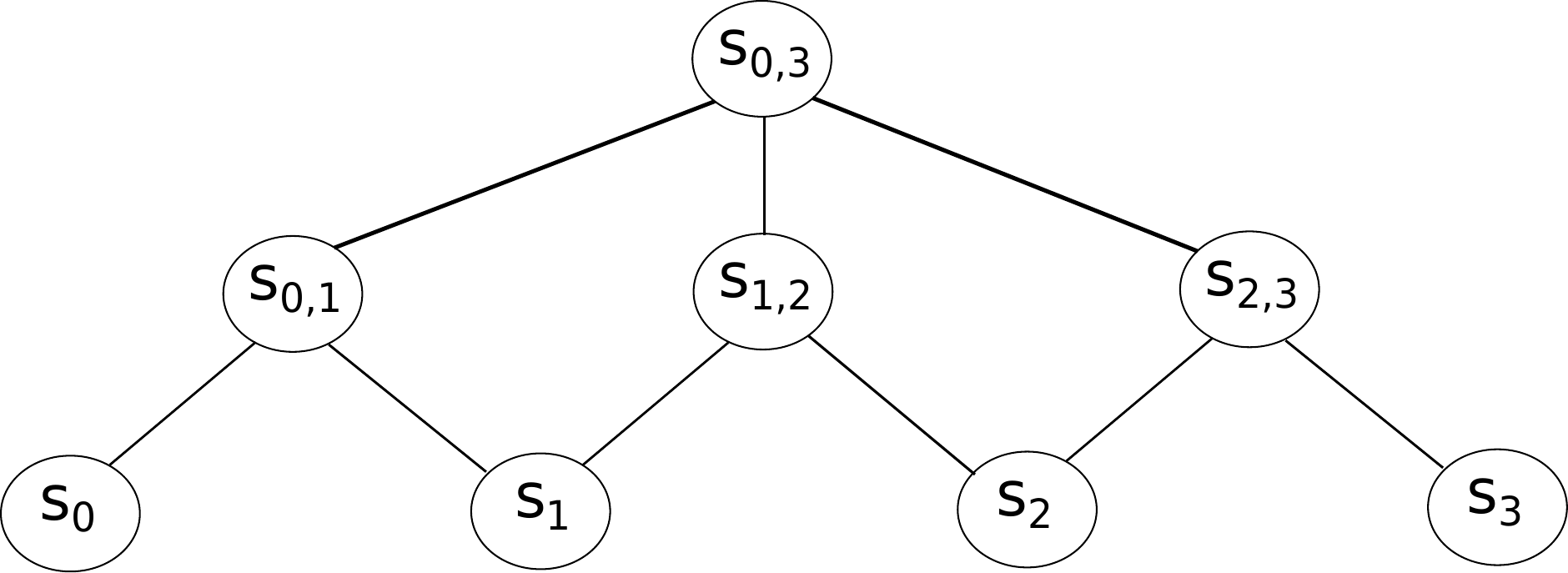}
\caption{Model for power-of-two requests. $T=4, p=2$}
\label{Hasse:power1}
\end{subfigure}
\vspace{-0.5cm}
\caption{Models of slot dependencies for a link.}
\vspace{-0.8cm}
\end{figure}

When a connection request is routed using a single slot, all slots in the spectrum, $s_0$ to $s_{T-1}$, 
can be used.
When the required number of slots is two, then two consecutive slots must be reserved, i.e., $s_0$ and $s_1$, or 
$s_1$ and $s_2$, etc. As the number of required slots grow, the dependencies between neighboring slots build up.
This relationship between the slots of a link can be modeled using a semi-lattice structure, where the 
set of all sets of neighbor slots form a partially ordered set. 
\begin{theorem}
Let $s_0, s_1, s_2, ..., s_{T-1}$ be the slots of the mini-grid on a fiber link. 
Let $k$ be the maximum number of consecutive slots required by the largest bandwidth request in the network.
Let $s_{i,j}$, where $i < j$ represent the set of all consecutive slots from $i$ to $j$: 
$\{s_{i}, s_{i+1}, s_{i+2}, ..., s_{j}\}$.
Then, the set
\begin{eqnarray*}
L = \{s_0, s_1, ... s_{T-1},
s_{0,1}, s_{1,2}, ..., s_{T-2,T-1},
s_{0,2}, s_{1,3}, s_{2,4},..., s_{T-3,T-1},
s_{0,3}, s_{1,4}, ..., s_{T-4,T-1},
...,
s_{0,k-1}, s_{1,k}, ..., s_{T-k, T-1}\}
\end{eqnarray*}
is a partially ordered set. Furthermore, $\wedge$ is a binary operation which is equivalent to set intersection, 
and $\vee$ is a binary operation equivalent to set union. 
Then the structure $< L, \wedge, \vee >$ is a join semi-lattice.
\end{theorem}
The Hasse Diagram of the semi-lattice defined can be used to model the interdependencies of the neighboring slots
in a link. An example diagram of the model with $T = 4, k = 4$ is given in Fig.~\ref{Hasse:uniform1}. Here, 
each node represents an element of set $L$, and an edge connects two nodes $s_{i,j}$ and $s_{l,m}$ 
if spectrum interval $[i,j]$ is a sub-interval of the spectrum interval $[l,m]$.
Each time a new slot set is assigned to a connection request, these slots and all the slot sets
including them become unavailable for future requests. This effect can be shown on the diagram in 
Fig.~\ref{Hasse:uniform1} by deleting the up-sets of all the assigned slots. For example, assigning $s_1$ to
a connection would cause the deletion of nodes $\{s_{0,1}, s_{1,2}, s_{0,2}, s_{1,3}, s_{0,3}\}$. In this case,
the diagram consisting of one connected component is separated into two connected components, one corresponding
to $\{s_0\}$, and the other corresponding to $\{s_2, s_3, s_{2,3}\}$. This effect also represents 
the fragmentation resulting from the assignment of $s_1$ to a connection.
\begin{theorem}
The total number of elements (nodes) in the join semi-lattice $<L, \wedge, \vee>$ representing the single link of an
elastic optical network, where the total number of slots is $T$ and the bandwidth requests are uniform, is
$N = k \left[ T - \frac{k - 1}{2} \right]$.
\end{theorem}

\subsection{Power-of-two requests}
Similar to the uniform requests case, the dependencies between the slots of a link can be modeled again using a 
lattice structure, where the partially ordered set is the set of all sets of neighbor slots where the number
of elements in a set is a power of two.
\begin{theorem}
Let $s_0, s_1, s_2, ..., s_{T-1}$ be the slots of the mini-grid on a fiber link. 
Let $2^p$ be the maximum number of consecutive slots required by the largest bandwidth request in the network.
Let $s_{i,j}$, where $i < j$ represent the set of all consecutive slots from $i$ to $j$: 
$\{s_{i}, s_{i+1}, s_{i+2}, ..., s_{j}\}$
Then, the set
\begin{eqnarray*}
L = \{s_0, s_1, ... s_{T-1},
s_{0,1}, s_{1,2}, ..., s_{T-2,T-1},
s_{0,3}, s_{1,4}, ..., s_{T-4,T-1},
...,
s_{0,2^{p}-1}, s_{1,2^{p}}, s_{2,2^{p}+1}, ..., s_{T-2^{p}, T-1}\}
\end{eqnarray*}
is a partially ordered set. Furthermore, $\wedge$ is a binary operation which is equivalent to set intersection, 
and $\vee$ is a binary operation equivalent to set union. Then the structure $< L, \wedge, \vee >$ is a join semi-lattice.
\end{theorem}
An example model of slot interdependencies for $T = 4, p = 2$ is given in Fig.~\ref{Hasse:power1}.
\begin{theorem}
The total number of elements in the join semi-lattice $<L, \wedge, \vee>$ representing the single link of an
elastic optical network, where the total number of slots is $T$ and the bandwidth requests are power of two is
$N = T (p + 1) - 2^{p+1} + p + 2$.
\end{theorem}

\section{Modeling the network}
\begin{figure}[t]
\centering
\begin{subfigure}{0.10\textwidth}
\centering
\includegraphics[width=2.5cm]{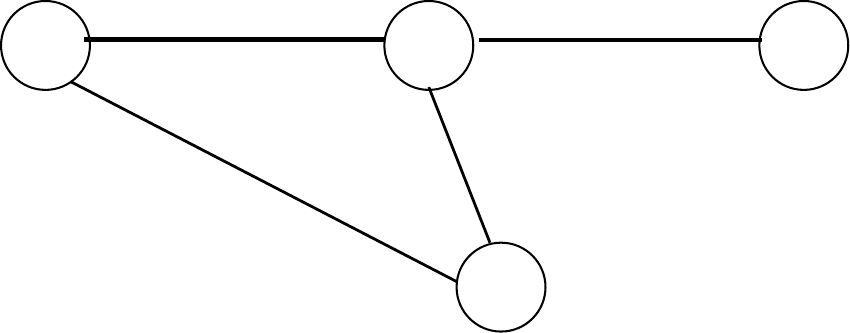}
\caption{A 4-node mesh topology.}
\label{network:4node}
\end{subfigure}
\begin{subfigure}{0.44\textwidth}
\centering
\includegraphics[width=7cm]{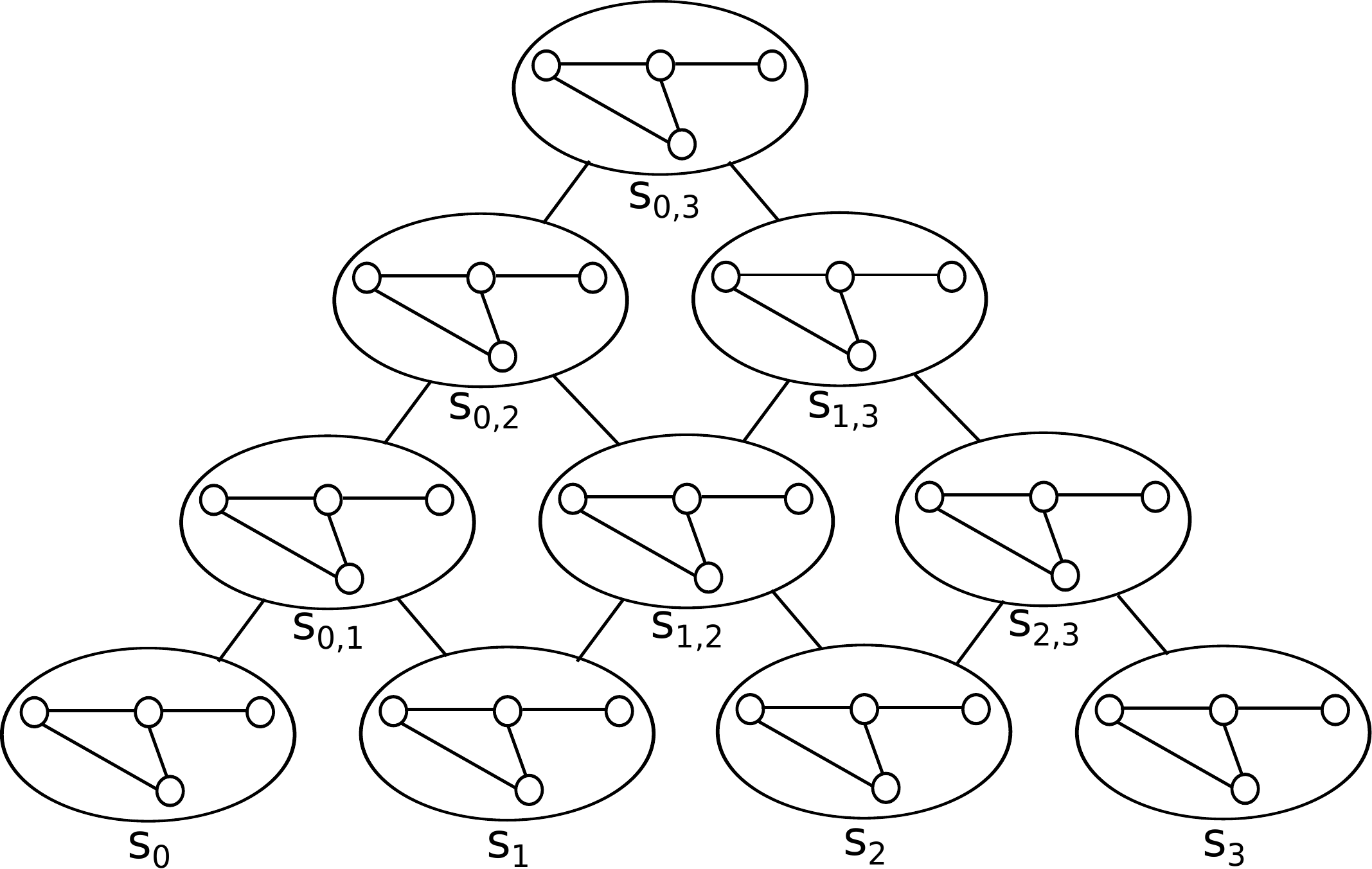}
\caption{Model representing the 4-node network.}
\label{network:model}
\end{subfigure}
\begin{subfigure}{0.44\textwidth}
\centering
\includegraphics[width=7cm]{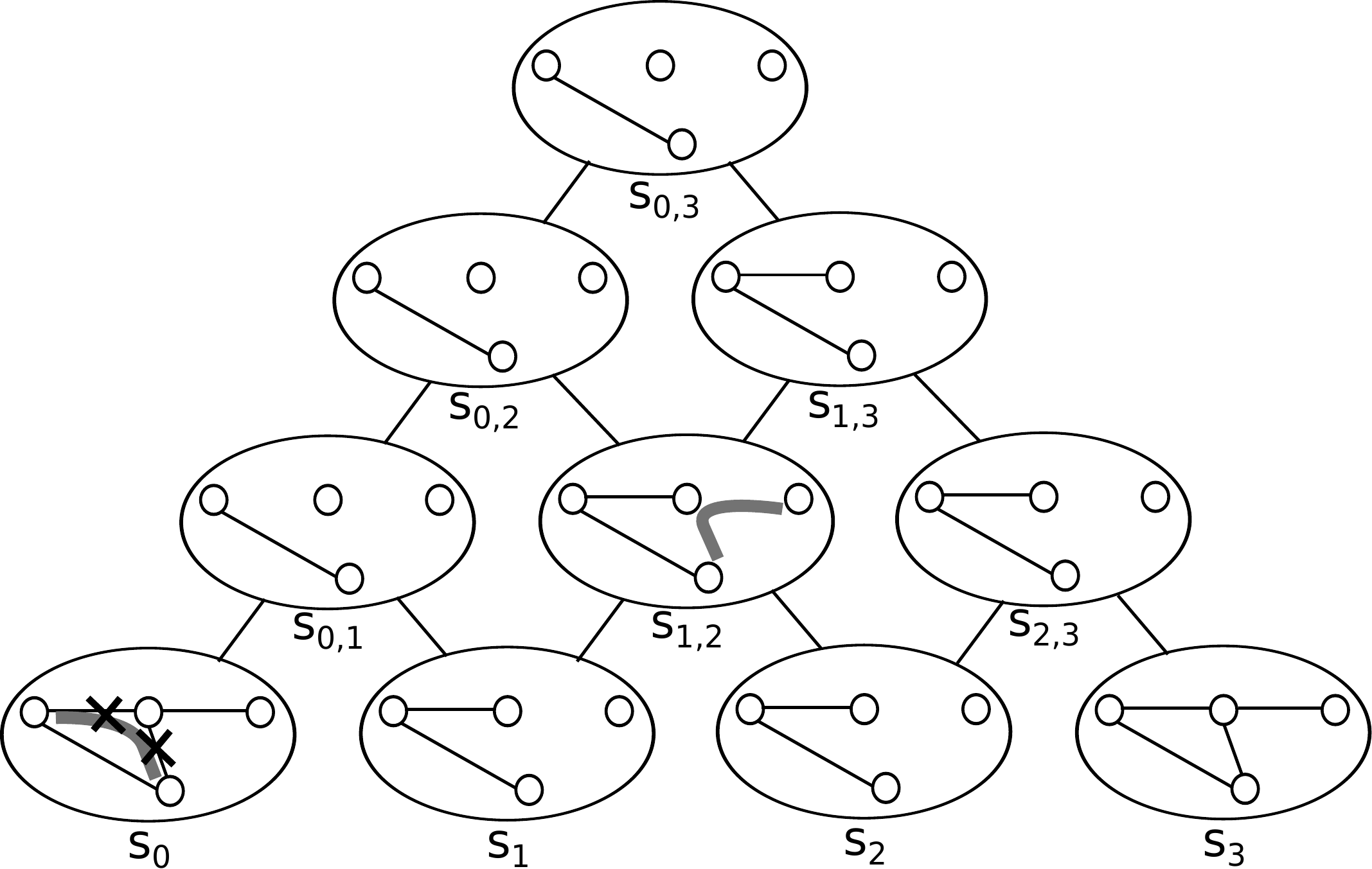}
\caption{Using the model: Set up of two connections.}
\label{network:rsa_ex}
\end{subfigure}
\vspace{-0.5cm}
\caption{Generalized layered model of a network.}
\vspace{-0.8cm}
\end{figure}

When we want to model the route and slot assignments of a whole network, the link representation 
described in the previous section can be used to form a new layered model to represent the spectrum
slot usage of a flexible-grid network.
In this model, each node represents a specific group of slots in the whole network 
topology. The groups including 1 slot are represented on the lowest part of the model graph which we call
level 1. In general, the groups including $i$ slots are represented by the nodes on level $i$.
The enumeration of node levels are between $1$ and $k$ for the uniform requests model. 
Assuming a simple mesh topology shown in Fig.~\ref{network:4node} for simplicity,
the proposed network model is given in Fig.~\ref{network:model}. In this example, 
we use the uniform request model, however, same model can be applied to any
traffic request pattern (e.g., power-of-two pattern). The level enumeration should be accordingly changed, e.g.,
for the power-of-two model the level numbers in Fig.\ref{Hasse:power1} are $1, 2$ and $4$ from bottom to top. 

\section{Using the model for routing and spectrum assignment}
The model in Fig.\ref{network:model} can be used for both static and dynamic routing and spectrum assignment
to represent the current status of spectrum usage in all the links. Having this representation readily will
cut down the complexity of the RSA algorithms.

An example usage of the model is given in Fig.\ref{network:rsa_ex} where two connections are set up.
In this example, we assume that first-fit spectrum assignment is used.
First, assume that there are no connections in the network, and the first request arrives between west and south
nodes with a bandwidth 
requirement of one slot ($b=1$). Then, a (shortest) path should be searched on the network graph at model node 
$s_{0}$ first, and if
we cannot find a path, we continue searching for a path on $s_{1}$, $s_{2}$, etc. Since in our example the
network is empty, we can find a path on $s_{0}$, and this path is shown on Fig.\ref{network:rsa_ex} as a gray
curve. Two links used to route this connection are deleted from the network of $s_{0}$ (This operation is 
explicitly shown on the figure with 'X' only for this specific node). 
Once the connection is set up, the slot $s_{0}$ becomes unavailable
for the up-set of $s_{0}$, i.e., $s_{0,1}, s_{0,2}, and s_{0,3}$. Therefore, same links are also removed 
from the networks at the respective nodes. Next,
assume that a second connection request between south and east nodes with $b=2$ arrives.
This time, we should search for a path on level 2 nodes, starting from the leftmost node (because of 
first-fit). The routing algorithm cannot find a path on node $s_{0,1}$ because the two links on the path of
the first connection are already removed. A path can be found on node $s_{1,2}$, which is shown on the figure
as a gray curve. Then the connection can be set up on spectrum slots $1$ and $2$, and the route links should
be removed from the sets and the up-sets of $s_{1}$ and $s_{2}$. 
The figure shows the final status of the network after the set up of these two connections.

When a connection is released, the links on the used slots and their respective up-set nodes should be restored.
This operation can be easily done by a logical \texttt{and} operation among the nodes of the model. 
In general, a link on a level $i$ node is 
available, if the same link is available on its down-neighbor nodes (connected by an edge) on level $i-1$.
This operation can be used both to remove and restore the links after a connection set up or 
tear down respectively. 

The model reduces the complexity of the RSA algorithms by a factor of $b$ (number of slots required by the 
connection), since slot-by-slot availability check is not required at the time of connection routing.

\section{Conclusions}
We proposed a new layered model for representing the spectrum assignment of flexible-grid optical networks, as the 
conventional layered representation designed for WDM networks cannot be used, due to slot interdependencies. 
The model can reduce the time complexity of the routing and spectrum assignment algorithms by a factor of $b$ 
(number of slots required per connection), and it is practical in terms of memory space requirements, even for
realistic networks.

\end{document}